\documentstyle[preprint,aps]{revtex}
\def\slashchar#1{\setbox0=\hbox{$#1$}           
   \dimen0=\wd0                                 
   \setbox1=\hbox{/} \dimen1=\wd1               
   \ifdim\dimen0>\dimen1                        
      \rlap{\hbox to \dimen0{\hfil/\hfil}}      
      #1                                        
   \else                                        
      \rlap{\hbox to \dimen1{\hfil$#1$\hfil}}   
      /                                         
   \fi}                                         %
\begin{document}
\draft
\title{Possible Detection of a Higgs Boson at Higher Luminosity Hadron
Colliders}
\author{Stephen Mrenna}
\address{California Institute of Technology \\
        Pasadena, CA}
\author{G.L. Kane}
\address{Randall Physics Laboratory, Univ. of Michigan \\
        Ann Arbor, MI }
\date{12 June 1994}
\maketitle
\begin{abstract}
   We have examined how a Standard Model or Supersymmetric Higgs boson
$h$ might be detected at possible hadron colliders.
The channels $W(\rightarrow\ell\nu)h(\rightarrow b\bar{b})$,
$Z(\rightarrow\ell\bar{\ell})h(\rightarrow b\bar{b})$ and
$W,Z(\rightarrow q\bar{q})h(\rightarrow\tau^{+}\tau^{-})$ are most useful.  The
results imply that $h$
with mass $M_h$
can be detected or excluded for 80~GeV$~{\stackrel{\scriptstyle <}{\scriptstyle
\sim}}~ M_h ~{\stackrel{\scriptstyle <}{\scriptstyle \sim}}~$130~GeV
at any hadron collider
with energy $\sqrt{\rm s~}$ $~{\stackrel{\scriptscriptstyle
>}{\scriptscriptstyle \sim}}~$2~TeV and an integrated luminosity
${\cal L} ~{\stackrel{\scriptscriptstyle >}{\scriptscriptstyle \sim}}~ 10$
fb$^{-1}$; high luminosity is the essential requirement.
For ${\cal L} ~{\stackrel{\scriptscriptstyle >}{\scriptscriptstyle \sim}}~ 30$
fb$^{-1}$, the $M_h$ reach is expanded beyond 130 GeV.
A \mbox{p--$\overline{\rm p}$~} collider is slightly better than a p--p
collider of
equal $\sqrt{\rm s~}$ and ${\cal L}$ for Higgs
detection.
We comment on measuring $h$ couplings and branching
ratios.
\end{abstract}
\pacs{}
%
%

\section{\rm Introduction}
Understanding the physics of the Higgs sector is the central task of
particle physics today.  This is necessary to complete the Standard Model
(SM), and the result will help point the way to extending the SM and
strengthening its foundations.  Probably it will be necessary to detect
or exclude a Higgs boson to make progress.

Discussing a Higgs boson in the SM is subtle, because no way is known
to maintain a light Higgs boson if any high scale exists.
If the
apparent perturbative unification of the SM gauge couplings \cite{langacker}
at a scale of the order $10^{16}$~GeV is not dismissed as a meaningless
accident, then there is an upper limit of about 200~GeV on the mass of the
SM Higgs boson
$h_{SM}$ \cite{cabi}.  In any supersymmetric theory, this upper limit
drops to about 150~GeV \cite{espinosa}.  These arguments are general.
While such indirect reasoning may not satisfy everyone, they surely imply
that the highest priority should be to search for the Higgs boson in the
mass range below about 200~GeV.  Its discovery will indicate that
supersymmetry (SUSY)
is probably realized in Nature, while its exclusion will be the final chapter
of low energy SUSY.

To put it succinctly, the best arguments known
today imply that the region $M_h =$~60--200~GeV
is the most important one,
and we have focussed on the most
difficult part: 80~GeV$~{\stackrel{\scriptstyle <}{\scriptstyle \sim}}~ M_h
{}~{\stackrel{\scriptstyle <}{\scriptstyle \sim}}~$130~GeV.
The region below 80~GeV will be covered at
LEP2, while the region above 145~GeV is rather easily studied at a hadron
collider with enough energy and luminosity in any case.  For
130--145~GeV, the situation needs to be analyzed in detail for a given
collider; we think this region can also be covered.
The remaining mass region, covering the gap from LEP2 up to about 125
GeV, is generally thought to be coverable by the rare two photon decay
of $h$ at the CERN LHC.  This requires an LHC electromagnetic calorimeter
with excellent energy resolution and pointing capabilities that is,
additionally, radiation hard.

We find that the inclusive production p + p$(\bar{\rm p})\rightarrow h + X$ is
not helpful to detect or exclude $h$, because of
a combination of small cross sections and large QCD backgrounds,
for hadron colliders with $\sqrt{\rm s~}$ $~{\stackrel{\scriptstyle
<}{\scriptstyle \sim}}~$14~TeV and
instantaneous luminosity $L < 10^{34}$cm$^{-2}$s$^{-1}$.
However, the
associated production processes $W+h$ and $Z+h$ can provide a signal for $h$
in several decay channels: $W(\rightarrow\ell\nu)h(\rightarrow b\bar{b}),
Z(\rightarrow\ell\bar{\ell})h(\rightarrow b\bar{b})$ and
$W,Z(\rightarrow q\bar{q})h(\rightarrow\tau^{+}\tau^{-})$.  The channels with
$W,Z\rightarrow $quark jets and
$h\rightarrow\tau^{+}\tau^{-}$ have not been considered elsewhere to our
knowledge.

We have examined many backgrounds to these processes, and present most of
our
results in terms of the number of signal events $S$ divided by the square
root of the number of background events $B$, $S/\sqrt{B}$.  We think it
is appropriate to combine channels, so the final results are the combined
significance for the four channels listed above.  Somewhat surprisingly,
we find that the behavior of signal and background with collider energy are
such that there is little or no gain in going to higher energies.  Rather,
luminosity is the key variable for $M_h ~{\stackrel{\scriptstyle
<}{\scriptstyle \sim}}~ 130$ GeV.

We find that,
with an instantaneous luminosity $L$
of order $10^{33} {\rm cm}^{-2}{\rm s}^{-1}$,
it is possible to detect or exclude $h_{SM}$ in the
region 80~GeV$~{\stackrel{\scriptstyle <}{\scriptstyle \sim}}~ M_{h_{SM}}
{}~{\stackrel{\scriptstyle <}{\scriptstyle \sim}}~ 130$~GeV at a hadron
collider
with $\sqrt{\rm s~}$ $~{\stackrel{\scriptscriptstyle >}{\scriptscriptstyle
\sim}}~$2~TeV.
Studies of supersymmetric theories \cite{kane2} have shown that if the
combined constraints of unification of the couplings, electroweak symmetry
breaking and consistency with data are imposed, then the lightest
SUSY Higgs boson
behaves essentially identically to a light
SM Higgs in all ways, so our results
apply equally as well to the lightest SUSY Higgs boson of the constrained
models.  In this paper, we show results for a 2 TeV \mbox{p--$\overline{\rm
p}$~}, 4 TeV p--p, and
a 14 TeV p--p collider with $L ~{\stackrel{\scriptscriptstyle
>}{\scriptscriptstyle \sim}}~ 10^{33}{\rm cm}^{-2}{\rm s}^{-1}$.

We assume a typical SSC or LHC detector is available to detect $h$.  The
detailed properties used in the analysis are described in Section 2.  The
most difficult requirement is that the detector must operate efficiently
at luminosities of order $10^{33}{\rm cm}^{-2}{\rm s}^{-1}$.
Sections 3--5 present the results for different channels, section 6 the
combined results, section 7 comments on measuring
$h$ properties, and section 8  some concluding remarks.

While we were preparing this paper, two other papers \cite{marciano,gunion}
appeared on
the same topic.  Where we overlap, the cross sections and
results are consistent.  We have considered
in addition the $W,Z(\rightarrow q\bar{q})h(\rightarrow\tau^{+}\tau^{-})$
channels,
so our conclusions are more favorable for Higgs detection.


\section{\rm Event Simulation}

   Calculations are parton level, unless a full simulation is needed for
a given signal or background, based
on the PAPAGENO \cite{papageno} and PYTHIA \cite{pythia} generators.
We use $m_t = 150$~GeV, 2nd order
running of $\alpha_s$ and CTEQ-1M structure functions.
The signal and backgrounds to $\tau^{+}\tau^{-}$ + jets
(such as $W^\pm\rightarrow\tau^{\pm}\nu_\tau +$ jets, with
a real $\tau$ and a jet that fakes the hadronic decay of
a $\tau$ in the final state)
were estimated using PYTHIA with initial and final state showering and
fragmentation.
We use no ``K-factors'' for the signals or backgrounds in the
individual channels considered, so our results will
improve when radiative corrections are included.  However, when presenting
the combined significance of all channels in Figure 2,
the effect of varying K over the
region $1 ~{\stackrel{\scriptstyle <}{\scriptstyle \sim}}~ {\rm K}
{}~{\stackrel{\scriptstyle <}{\scriptstyle \sim}}~ 1.2$ is illustrated by a
light shaded band.

   The following detector properties are implicit to this analysis:
\begin{itemize}
\item   Hadron calorimetry resolution $60\%/\sqrt{\rm E }\oplus 4\%$ covering
$|\eta| \le$ 2.5.
\item   Electromagnetic calorimetry resolution $7\%/\sqrt{\rm E}\oplus .5\%$
covering $|\eta| \le$ 2.5.
\item   Muon acceptance and momentum resolution comparable to electrons.
\item   Forward calorimetry covering to $|\eta|=$ 5 and
   $\slashchar{E}_{t}$
   resolution 40\%/$\sqrt{\slashchar{E}_t}$.
\item   A central tracker with good impact resolution and a high reconstruction
   efficiency to allow the tagging of $b$--jets and $\tau$ decays.
\end{itemize}

The choice of hadronic calorimetry is slightly worse than that
available presently at the ${\bf D0}$ detector at FNAL, but is
slightly better than that at  ${\bf CDF}$.
The excellent electromagnetic energy resolution was chosen for optimizing
the $h\rightarrow\gamma\gamma$ and $h\rightarrow Z^{(*)}Z^{(*)}$ detection
studies.
Since these channels were found not to be significant with the Higgs mass
range and integrated luminosity considered, such electromagnetic
resolution is not essential to the remainder of this study.


\section{$Z(\rightarrow\ell\bar{\ell})h(\rightarrow {b\bar{b}})$}

The signal considered is $Z + h, Z\rightarrow \nu\bar{\nu}, e^+e^-,$ or
$\mu^+\mu^-$,
and $h\rightarrow b\bar{b}$.  In the SM, $BR(h\rightarrow b\bar{b})$ = .83
(.65) for $M_h$ =
80 (120) GeV.   The $Z$ boson decays to the lepton final states
$\nu\bar{\nu},e^+e^-,$ and $\mu^+\mu^-$
with a branching ratio $\approx 27\%$.

The kinematic cuts requested for the leptons are:
\begin{itemize}
\item $p_{t}^{\ell} >$ 20~GeV, $|\eta^{\ell}| <$ 2.5,
 and~ $ |M_{\ell\bar{\ell}}-M_Z | <$ 5 GeV, for
$Z\rightarrow\ell\bar{\ell}, \ell=e,\mu,$ and
\item $\slashchar{E}_{t} >$ 20~GeV, for $Z\rightarrow\nu\overline{\nu}$.
\end{itemize}

We also require the reconstruction of 2 jets $j$,
where $j$ is defined in the standard manner
with R~=~.6 as the jet cone size, such that:

\begin{itemize}
\item $E_{t}^{j} >$ 20~GeV and $|\eta^{j}| <$ 2.5.
\end{itemize}

In addition, one can demand a $|\cos\theta^*|$ distribution for the
jets, where
$\theta^*$ is the decay angle of the jets in the jet--jet rest frame,
that is consistent with higgs decay.  We find that, after the previous cuts,
such an additional cut is not effective and only reduces the significance
of the signal.

Each jet $j$ is required to pass a single or double heavy flavor tag, where
the efficiencies for tagging $b$--jets, $c$--jets, and $g,u,d,s$--jets as
$b$--jets
are (${\epsilon}_{b}^{b},{\epsilon}^{b}_{c},{\epsilon}^{b}_{j}$).
We considered several sets of tagging efficiencies for both tagging
scenarios.
In this paper, we only report results for one scenario, (40\%,5\%,1\%).
We think this number is quite reasonable, since the
optimal tagging efficiency will result from a combination of
impact parameter, soft lepton, multivariate regression analysis,
and other techniques.
Recently, CDF estimated their b-tagging efficiency to be ${\epsilon}_{b}^{b}
\approx$ .22
\cite{CDFTop}.

The backgrounds considered are:

\begin{itemize}
\item $Z(\rightarrow\ell\bar{\ell}) j_1 j_2$, where $j_1$,$j_2$ are any
combination of $g,u,d,s,c,$ or $b$ (and their anti-particles),
\item $Z(\rightarrow\ell\bar{\ell})Z(\rightarrow b\bar{b})$, and
\item $t\bar{t}\rightarrow bW^+ \bar{b}W^-$.
\end{itemize}
The $W$-bosons produced in $t$--decay are allowed to decay to leptons or jets.

We emphasize that the $Zj_1 j_2$ background includes $Zbj$ and
$Zcj$ (where $j$ is specifically $g,u,d,s$ and their anti-particles)
final states which are single tagged with efficiency $\simeq
{\epsilon}_{b}^{b}$ and
${\epsilon}^{b}_{c}$ and double tagged with an efficiency
${\epsilon}_{b}^{b}\times{\epsilon}^{b}_{j}$ and
${\epsilon}^{b}_{c}\times{\epsilon}^{b}_{j}$, respectively.
In Table 1(a) we list the signal and background for various hadron colliders,
using a single $b$--tag at a 2 TeV \mbox{p--$\overline{\rm p}$~} collider and a
double $b$--tag for
a 4 TeV and 14 TeV p--p collider, assuming 30 fb$^{-1}$ of integrated
luminosity.  The cross sections listed are those after all
cuts and $b$--tag requirements.  We also show the Gaussian width of the
reconstructed signal and the significance.

Once a candidate bump is detected in the invariant mass spectrum of
the reconstructed $b\bar{b}$ pair,
the ``signal'' is the number of excess events over a smooth --
relatively flat -- background
within $\pm 2~ \sigma_M$ of the central value of the bump, where $\sigma_M$ is
the Gaussian width of the observed signal.
For $M_h =$ 80 (120) GeV, $\sigma_M$
$\simeq$ 5.5 (7.2) GeV.

As pointed out by Gunion and Han \cite{gunion},
the semi--leptonic decays of the $b$-- and
$c$--quarks affect the reconstructed invariant mass of the $b$-quark pair
from $h$ decay.  In addition, final state showering will have an affect.
{}From a full analysis of the fragmentation and decays of $h\rightarrow
b\bar{b}$ using
JETSET 7.4 \cite{jetset}, we find that the peak of the invariant mass
distribution is nearly the
same as for $h\rightarrow{q}\bar{q}$, where $q$ is a light quark,
but the shape is skewed to lower values.
The average of the distribution is typically \mbox{3--4~GeV} below the peak.
We feel that a $b$--jet reconstruction algorithm can be developed based on
our knowledge of the $B$--meson mass and lifetime, the $b$--quark fragmentation
function, the kinematics of semi--leptonic decays and a measurement of
the $\slashchar{E_t}$.  If uncorrected, we find that tighter cuts on the
non--Gaussian invariant mass distribution retain a signal with a significance
reduced by 8--10\%.


\section{$W(\rightarrow\ell\nu)h(\rightarrow {b\bar{b}})$}

The signal considered is $W + h, W\rightarrow e\nu_e$ or $\mu\nu_\mu$,
and $h\rightarrow b\bar{b}$.
The $W$ boson decays to $e\nu_e$ and $\mu\nu_{\mu}$
with a branching ratio $\approx 2/9$ = .22\%.

The kinematic cuts requested for the leptons are:
\begin{itemize}
\item $p_{t}^{\ell} >$ 20~GeV and $|\eta^{\ell}| <$ 2.5, $\ell = e,\mu$, and
\item $\slashchar{E}_{t} >$ 20~GeV.
\end{itemize}

The same jet and tagging requirements are used as previously stated.
The backgrounds considered are:
\begin{itemize}
\item $W^{\pm}(\rightarrow\ell^{\pm}\nu)j_1 j_2$,
where $j_1 j_2$ are any combination of $g,u,d,s,c,$ or $b$ (and their
anti--particles),
\item $W^{\pm}(\rightarrow\ell^{\pm}\nu)Z(\rightarrow b\bar{b})$,
\item $t\bar{t}\rightarrow bW^+\bar{b}W^-$, and
\item
$q\overline{q}^{'}\rightarrow W^{*}\rightarrow t\bar{b}\rightarrow
b\bar{b}W^{\pm}(\rightarrow\ell^{\pm}\nu)$.
\end{itemize}

The $W$--bosons from $t$--decay are allowed to decay to leptons or jets.

In Table 1(b), we summarize the results for various hadron colliders in the
same manner as for the $Z(\rightarrow\ell\bar{\ell})h(\rightarrow b\bar{b})$
channel.
In Figure 1(a), we show a simulation of the signal and the
background, with the background subtracted, in this channel for
$M_h \simeq M_Z$ at a 2 TeV \mbox{p--$\overline{\rm p}$~} collider with an
integrated luminosity
of 30 fb$^{-1}$.  In this figure, we have applied a single $b$-tag to
all events.


\section{$W/Z(\rightarrow jj)h(\rightarrow\tau^{+}\tau^{-})$}

The final signal considered here is $W/Z + h, W/Z\rightarrow$ jets,
$h\rightarrow\tau^{+}\tau^{-}$.
In the SM, the $BR(h\rightarrow\tau^{+}\tau^{-}$) is
.080 (.067) for $M_h$ = 80 (120) GeV.
For this channel, we do not use the leptonic decays of the
heavy gauge bosons to reduce backgrounds, but, rather, the fact that
$\tau$ decays have a low multiplicity of secondaries and can
produce a large $\slashchar{E_t}$.  The dominant $\tau$
decay modes are:
\begin{itemize}
   \item $BR(\tau^{\pm}\rightarrow\ell^{\pm}\nu_\ell\nu_\tau) \simeq 35\%, \ell
= e,\mu$, and
   \item $BR(\tau^{\pm}{\rightarrow}h^{\pm} + \ge 0$~ neutrals $)\simeq 50\%$,
\end{itemize}
where $h^{\pm}$ includes $\pi^{\pm}, \rho^{\pm}$, and $K^{\pm}$.
The remaining significant decay modes contain 3 charged pions and zero or more
neutral particles.

Since the $\tau$ pairs produced from $h$ decay are extremely energetic
compared to $m_\tau$ and have low multiplicities,
the decay products of each $\tau$ travel nearly in the same
direction as the parent $\tau$.
Therefore,
the measurement of the momentum of the secondary charged track determines,
to high accuracy, the
primary $\tau$'s direction of motion.  The measured
$\slashchar{E_t}$ vector, projected onto the charged track's direction,
determines the full $p_t$ of the $\tau$.
Given the $p_t$ of the $\tau$ and assuming all the decay products point
in the direction of the observed charged track (which is also, by assumption,
the moving direction of the $\tau$), we fully reconstruct the $\tau$
momentum.  The reconstruction works much better if the $\tau$ pair system
is boosted, thereby removing the possibility that the $\slashchar{E_t}$ of
the neutrinos totally destructively interfere.

Once the momentum of the
$\tau$'s has been reconstructed in this method, the validity of the
approximation can be tested.  Taking, for example,
the case $\tau^{\pm}\rightarrow\pi^{\pm}\nu_\tau$,
the goodness of the reconstruction is tested by evaluating
$\cos\delta = 1-\frac{m_\tau^2}{E_{\pi}E_{\nu}}$ for each $\tau$, where
$E_\nu$ and $E_\pi$ are the energies of the reconstructed
neutrino and measured charged track,
respectively.  The $\tau$ with the smallest value of $\cos\delta$ has been
reconstructed more poorly than the other.  For this $\tau$, the
$\slashchar{E_t}$ should not be projected onto the charged track's
direction of motion, but, rather, onto a vector lying on a cone with
opening angle $\delta$ with respect to the charged track.  By proceeding
in this manner and choosing the neutrino direction of motion on this
cone that minimizes
the invariant mass of the $\tau$ pair, a better estimate is made of the
invariant mass.  This method is just a crude attempt at obtaining a
better measurement of $m_{\tau^{+}\tau^{-}}$, and a more detailed algorithm
will do better.

Since a good measurement of $\slashchar{E_t}$ is needed, we do not consider the
leptonic decays of the $W$ or the neutrino channels of the $Z$, but, instead,
reconstruct the $W$ or $Z$ in jets.  Fortunately, this still leads to a
detectable signal.
Concentrating on the one-prong final states of the $\tau$, there are
$\tau^{+}\tau^{-}$ final states $\ell^{+}\ell^{-}$, $\ell^{+}\pi^{-}$,
$\pi^{+}\ell^{-}$, and $\pi^{+}\pi^{-}$.  Here, and in the
following discussion, $\ell$ refers to $e$ or $\mu$, while $\pi$ includes
$\rho$ and $K$ mesons.
The previously discussed channels all contained a high $p_t$
charged lepton, $\slashchar{E_t}$, or combination of these to use as a
trigger.  It is reasonable that a combination of $\slashchar{E_t}$ and
the isolation of the $\pi$ from neighboring hadronic activity can be used to
trigger on the $\pi^+\pi^-$ final states.  We considered two cases, without and
with an isolated $\pi$ trigger.  The $\ell^+\ell^-$, $\ell^+\pi^-$,
and $\pi^+\ell^-$ final states
have a combined branching ratio of .47, while including the $\pi^+\pi^-$
final states increases this to .72.

Each event must contain at least 2 jets $j$,
defined with R = .6, and 2 reconstructed
$\tau$'s
satisfying the
following requirements:
\begin{itemize}
\item $E^j_t >$ 15 GeV, $|\eta^j| <$ 2.5,
\item $M_{W}  - 15$ GeV $< M_{jj} < M_{Z} + 15$ GeV,
\item $p_{t}^{\tau} >$ 20~GeV and $|\eta^{\tau}| <$ 2.5,
\item $\slashchar{E}_{t} >$ 20~GeV, and
\item  $m^{(lo)}_T>$ 20~GeV, $m^{(hi)}_T>$ 40~GeV,
where $m_T$ is the transverse
mass of the leptons or hadrons from the $\tau$ decays and $\slashchar{E_t}$,
and $(lo)$ and $(hi)$ refer to the smaller and larger values of $m_T$.
\end{itemize}

The one-prong decays of the $\tau$ are reconstructed
by finding a
charged track $(t)$,
with $p^{(t)}_t >$ 5 GeV, $|\eta^{(t)}|<$ 2.5,
pointing to the center of a narrow calorimeter
bump, so that
$\Delta R_{(t)j} \le $ .15.
$m_{\tau^{+}\tau^{-}}$ is reconstructed by projecting
the measured $\slashchar{E}_{t}$
onto the two charged tracks, as explained previously,
keeping only those events
that give a positive magnitude for the reconstructed $p^{\tau}_t$.

The backgrounds are events with
two leptons or one lepton and a jet that fakes the hadronic decay of a $\tau$,
a large $\slashchar{E}_{t}$, and a combination of jets that give an invariant
mass near the $W$ or $Z$ mass.
The jet that fakes the hadronic decay of a $\tau$
must not only have a single charged
track pointing to the core of its calorimetric cluster, but also
a $\slashchar{E}_{t}$ component pointing in its direction.
We ignore the possibility of a large fake $\slashchar{E}_{t}$
measurement because this is kinematically limited by the beam
energy and $|\eta|$ coverage of the forward calorimeter.  Even if a jet
with the
beam energy were lost ``down the beam--pipe'',
the maximal $\slashchar{E_t} =
E_{\rm beam}\sin\theta \simeq \frac{1}{2}E_{\rm beam}\exp^{-|\eta|}$, for
small $\theta$ measured from the beam--pipe.  All the backgrounds considered,
then, have at least one neutrino from the decays of $W$'s or $Z$'s.

The backgrounds considered are:
\begin{itemize}
\item $W^{\pm}(\rightarrow jj)Z(\rightarrow\tau^{+}\tau^{-})$,
\item $Z(\rightarrow jj)Z(\rightarrow\tau^{+}\tau^{-})$,
\item $t\bar{t}\rightarrow\ell$ + jets or $\ell\ell^{'}$ + jets, where
$\ell,\ell^{'} =
e,\mu,\tau$,
\item $W^{\pm}(\rightarrow\ell^{\pm}\nu)$+jets, where $\ell=e,\mu,\tau$, and
\item $Z(\rightarrow\tau^{+}\tau^{-})jj$.
\end{itemize}

The probability that a jet is tagged as a $\tau$ is estimated,
very conservatively, at 5\%,
based on a particle--level
simulation including fragmentation.  This estimate relies only on
the charged track multiplicity, not on a shower--shape analysis, which
could significantly reduce this number \cite{atlas}.
The $t\bar{t}$ and $W^{\pm}$ + jets
backgrounds are estimated by first finding a candidate jet that can fake
the hadronic decay of the $\tau$, then multiplying the final event rate by
this probability.
The decay $b\rightarrow c\tau\nu_\tau$ is
included in the JETSET  decay tables, which were used in
estimating the jet misidentification probability for $b$--,$c$-- and
light quarks and gluons.

We considered the scenario where only the leptonic $\tau$ decays can be
used as a trigger ($\ell$ trigger) and where the single-charged track decays
can be used as well ($\pi$ trigger).  The results presented assume
an isolated $\pi$ trigger.  The loss of significance of the signal
in using only the
$\ell$ trigger is typically 35\%.
Because of the importance of the resonant background
$Z(\rightarrow\tau^{+}\tau^{-})jj$,
the significance of the signal is calculated using MINUIT \cite{minuit}, which
accounts for the shape of the signal and background.
We stress that this background will be
normalized to high accuracy by observed $e^+e^- jj$ and $\mu^+\mu^- jj$ events
and assuming lepton universality.
In Table 1(c), we summarize our significance results for various
hadron colliders
by presenting the number of signal events for 30 fb$^{-1}$ of data
and the error on this number as determined by a functional fit to
the signal and background.
The significance is the ratio of these numbers, which
correspond to
$S$ and $\sqrt{B}$ in the previous discussion of significance.
We do not present results
for $M_h \simeq M_Z$ or for a collider with $\sqrt{\rm s~}$ = 14 TeV, since,
for those cases, a significant signal could not be found in this
channel.
In Figure 1(b), we show a simulation of the signal and the
background, with the background subtracted, in this channel for
$M_h$ = 120~GeV at a 2 TeV \mbox{p--$\overline{\rm p}$~} collider with an
integrated luminosity
of 30 fb$^{-1}$.

Since $\tau$ signatures at hadron colliders have not been well-studied,
we suggest several improvements
which will increase the significance of
this channel once they are included:
\begin{itemize}
  \item A better estimate of jet rejection will eliminate
backgrounds to the hadronic $\tau$ decay modes.
  \item Incorporating the correct polarization for $\tau$ decays
      may be useful, since $\tau$'s from $h$ decay
      lead to a {\it soft-soft} and {\it hard-hard}
      momentum correlation between the
      two charged hadrons.
  \item An impact parameter cut would reduce drastically all but
      the real $\tau$ and
      heavy quark backgrounds.
  \item Inclusion of the 3-prong decays of the $\tau$ will increase the
      signal event rate by 28\%.  Jet backgrounds can be reduced by
      using the kinematic constraint $m_{\pi\pi\pi} < m_\tau$ for the
      three charged tracks $\pi$.
  \item Better mass resolution, based on a more sophisticated $\tau$
      reconstruction algorithm,  will reduce the contribution of
$Z(\rightarrow\tau^{+}\tau^{-})jj$.
  \item If all or some of the above improvements can be realized, the
      direct production process p + p$(\bar{\rm p})\rightarrow
h(\rightarrow\tau^{+}\tau^{-})+X$
\cite{hinchliffe}, which has
      a larger cross section but potentially more backgrounds, may be
      accessible.
\end{itemize}


\section{\rm Combined Significance}

The significance of a signal $S$ above a background $B$, $S/\sqrt{B}$ in
the case of large statistics for a signal bump on a flat background, is
a measure of the probability that the background has fluctuated up to
fake the signal.  The probability is the same as exceeding $S/\sqrt{B}$
standard deviations
of a Gaussian probability function.   When combining two signals, where the
probability that the background has fluctuated up to the signal is
$p_1$ and $p_2$, respectively, one has
several options:  (1) use $p_1{\times}p_2$ as the combined probability,
(2) use $\alpha(1-\ln\alpha)$, where $\alpha=p_1{\times}p_2$, as a statistic,
and determine the probability that a second measurement
\mbox{$\alpha^{'}=p_1^{'}\times{p_2^{'}} < \alpha$},
or (3) combine the two signals and backgrounds
as though coming from the same experiment.  Method (1) will reject the
hypothesis that the signal is consistent with background if $p_1$ is very
small, even if $p_2 \simeq 1$.  Method (2) compensates for this by sampling
all combinations of $p_1$ and $p_2$ that could lead to a given
$\alpha=p_1\times{p_2}$.  Method (3) treats the two measurements as
independent data sets which are used to test the hypothesis that the
signal is consistent with background in exactly the same way.
Methods (1) and (2) will differ significantly only in extreme cases, which
are not present in this analysis.
For the cases when all channels
yield a significance $\sim$ 5, we simply use method (3); otherwise,
we use method (1), since we desire a high
significance in only one channel.

Figure 2 shows the combined significance of all channels considered in this
study as a function of $M_h$.  Figure 2 contains three
graphs, one for each hadron collider option, and each graph contains two
separated bands, which are further divided by different shadings.
The upper bands show the significance for an integrated luminosity of
30~fb$^{-1}$, the lower bands for 10~fb$^{-1}$.
The dark shading shows the range of significance
from varying the $Q^2$ scale of the
Electroweak-QCD background processes from $\bar{m}_T^2$ to $\hat{s}/2$.
$\bar{m}_T^2$ is the square of the average transverse mass ${m_T}_j$
of the outgoing
partons $j$ of mass $m_j$,
${m_T}_j = \sqrt{{p_T}_j^2 + {m}_j^2}$, and $\hat{s}$ is the
invariant mass of the hard-scattering parton process.  The lower bound of
the dark band corresponds to $Q^2 = \bar{m}_T^2$, the upper bound to
$Q^2 = \hat{s}/2$.  For example, at a $\sqrt{\rm s~}$ = 2~TeV hadron collider
with
and integrated luminosity ${\cal L} = 30~$fb$^{-1}$, the significance of
the signal for $M_h = 100$~GeV varies from 7.3 to 8.3 by choosing
$Q^2 = \bar{m}_T^2$ or $\hat{s}/2$.
Also, when radiative corrections are applied, typically the shape of
kinematic distributions of a given process are
only slightly changed, but the
magnitude of the distribution is scaled by a ``K--factor''.
The light shading shows the range of significance
from multiplying the signal by a
``K-factor'' of 1 to 1.2.  Therefore, the absolute lower bound for
each band in Figure 2 corresponds to $Q^2 = \bar{m}_T^2$ and K=1,
the absolute upper bound to $Q^2 = \hat{s}/2$ and K=1.2.

Several comments concerning the results presented in
Figure 2 are in order.  First, the $Q^2$ dependence
of the $W/Zjj$ backgrounds, which arises in evaluating the structure functions
$f_i(x,Q^2)$
and $\alpha_s(Q^2)$, is more important at a lower energy collider \cite{han}.
Since the
kinematic cuts have been chosen to accept the signal efficiently, it is
expected that the background which passes these same cuts has a topology
similar to the signal with the Higgs resonance replaced by an off-shell gluon.
In this case, a choice of $Q^2 = \hat{s}/2$ is better motivated than
$\bar{m}_T^2$.  Secondly, the $W,Z(\rightarrow
q\bar{q})h(\rightarrow\tau^{+}\tau^{-})$ channel increases
the reach of the lower energy colliders.  The increase of the $Zjj$,
$t\bar{t}$,
and $W$ + jet backgrounds and the decrease in importance of the
p + p $\rightarrow W/Z + h$ processes make a signal in this channel
untenable at \mbox{$\sqrt{\rm s~}$ = 14 TeV}.
Of course, higher luminosity and consideration of p + p $\rightarrow
h(\rightarrow\tau^{+}\tau^{-}) + X$
might change this conclusion.
Finally, though for brevity not presented here,
we also
considered a 4 TeV \mbox{p--$\overline{\rm p}$~} hadron collider.
For the same integrated luminosity,
a 4 TeV \mbox{p--$\overline{\rm p}$~} collider has a significance $\sim$
20--30\% higher than
a 4 TeV p--p collider.


\section{\rm Studying Higgs Properties}

Ideally, it would be possible not only to detect a Higgs boson, but also to
study its properties, establish that it interacted like a Higgs boson, and
even distinguish $h_{SM}$ from $h_{SUSY}$.
Showing that the spin of a
detected $h$ is zero will not be hard.
The important observables at a hadron collider are of the form
$\sigma\times BR$.  The full width $\Gamma_h$, for example, is
probably too narrow
to overcome experimental resolution.  Using the decay modes
discussed in this paper,
the following set of equations can be written:

\[
\begin{array}{ccc}
{\cal{L}}\times\sigma_{Wh}\times BR_b      & = & N_b \\
{\cal{L}}\times\sigma_{Wh}\times BR_{\tau} & = & N_{\tau} \\
BR_b + BR_{\tau}            & \approx & 1, \\
\end{array}
\]
where $\sigma_{Wh}$ is the cross section for the associated production
of $h$ with $W$ or $Z$, $BR_b$, $BR_\tau$ are the branching ratios for
$h(\rightarrow b\bar{b})$ and $h(\rightarrow\tau^{+}\tau^{-})$, respectively,
and
${\cal{L}}$ is the integrated luminosity of the collider in units of
inverse cross section.
Solving these equations allows us to write ${\cal{L}}\times
\sigma_{Wh} = N_b + N_\tau$.
$N_i$ is the number of $i$ type events corrected for reconstruction
efficiencies, tagging efficiencies, etc.
{}From the number of observed events
at a hadron collider, it will be possible to measure $g^2_{WWh}\times{BR_b},
g^2_{ZZh}\times{BR_b}, g^2_{WWh}\times{BR_\tau}$,
and $g^2_{ZZh}\times{BR_\tau}$.
The ratio $BR_\tau/BR_b$ is predicted to be $\frac{m^2_\tau}{3 m^2_b(M_h)}
\simeq \frac{1}{9}$ for any gauge theory, since $b$ and $\tau$ are in the
same position in any doublet; thus its measurement could establish that a
detected
boson was indeed coupling proportional to mass, but could never distinguish
among theories.  Similarly, since
$BR_\mu/BR_\tau = m^2_\mu/m^2_\tau \simeq\frac{1}{300}$, no excess of
events should be seen in $\mu^+\mu^-$ final states.  Given $BR_b/BR_\tau$,
the equality of the $WWh$ and $ZZh$ couplings could be checked at a 2 TeV
hadron collider.

We have not found a way to measure the $t\bar{t}h$ or $c\bar{c}h$
coupling at a low energy hadron
collider; the first of them can be measured at LHC,
which
corresponds to the $\sqrt{\rm s~}$ = 14 TeV collider considered here but with
an instantaneous luminosity of $L = 10^{34}$cm$^{-2}$s$^{-1}$,
and NLC.  Measuring the
$t\bar{t}h$ or $c\bar{c}h$ coupling could, in principle, distinguish among
theories, since the relative coupling of $T_3 = \pm\frac{1}{2}$ states
{\it does} depend on the theory, though in constrained SUSY theories these
couplings are expected to be very close to their SM values.
The important $\gamma\gamma$ coupling, with $BR \simeq 10^{-3}$, can only be
measured at LHC, where the cross section and luminosity are both large,
or at a photon collider.


\section{\rm Conclusions}

Developing the ability to do $\tau$ physics at a hadron collider is a
natural extenstion of the present ability to do heavy quark physics, and
can have a significant impact on the ability to do Higgs physics at a
hadron collider.

We see from Tables 1(a)--(c) and Figure 2 that
the detection or exclusion of $h$ for $M_h ~{\stackrel{\scriptstyle
<}{\scriptstyle \sim}}~ 130$~GeV is not only
possible at a 2 TeV high luminosity \mbox{p--$\overline{\rm p}$~} collider
(10--30 fb$^{-1}$),
it is competitive with -- if not better than --
hadron colliders at a higher $\sqrt{\rm s~}$,
because of the relative behavior of signal and
backgrounds.
The $W,Z(\rightarrow{jj})h(\rightarrow\tau^{+}\tau^{-})$ channel is the
strongest far enough above
$M_h \simeq M_Z$; near $M_Z$, the $W(\rightarrow\ell\nu)h(\rightarrow
b\bar{b})$ and
$Z(\rightarrow\ell\ell)h(\rightarrow b\bar{b})$ channels are adequate.  It is
probably possible to
detect $h$ above 130~GeV, but more analysis is needed to establish that.
The ratio of $\tau^{+}\tau^{-}$ and $b\bar{b}$ branching ratios and the ratio
of the couplings
of $h$ to $WW$ and $ZZ$ can be measured at a 2 or 4 TeV hadron collider.
The $t\bar{t}h$ coupling and the $\gamma\gamma$ branching ratio will probably
only be accessible at LHC or NLC.

Our analysis shows that a \mbox{p--$\overline{\rm p}$~} collider with
$\sqrt{\rm s~}$ = 2~TeV and integrated
luminosity ${\cal L} ~{\stackrel{\scriptscriptstyle >}{\scriptscriptstyle
\sim}}~$ 10 fb$^{-1}$ can detect or exclude a Higgs boson
with a mass ranging from the present upper limit up to about 130 GeV (and
perhaps higher), thus covering the region of greatest interest.  This is
a remarkable physics opportunity.


%
%
\setcounter{table}{0}
\begin{table}[htb]
\caption{(a) Sensitivity for the
$Z(\rightarrow\ell\bar{\ell})h(\rightarrow b\bar{b})$ channel}
\begin{center}
\begin{tabular}{||l|c|c|c|c|c|c||}
\tableline\tableline
 & Mass (GeV) &  80 &  90 & 100 & 110 & 120 \\
   &   $\sigma_M$ (GeV)           & 5.5 & 6.0 & 6.5 & 7.0 & 7.4 \\ \tableline
Cross Section (fb) &
SM Higgs     &    23. &    18. &    14. &    11. &  7. \\
$\sqrt{\rm s~}$ = 2 TeV, \mbox{p--$\overline{\rm p}$~} &
Backgrounds   &   242. &   246. &   212. &   176. &   152.   \\ \cline{1-7}
S/$\sqrt{\rm B}$ for 30 fb$^{-1}$   &
      1 Tag  &   8.0 &   6.2 &   5.0 &   4.2  &  3.0   \\
\tableline\tableline
%
%
Cross Section (fb) &
SM Higgs     &    9.1 &    7.0 &    5.2 &    3.8 &   2.6  \\
$\sqrt{\rm s~}$ = 4 TeV, p--p  &
Backgrounds     &  56. &  62. &   52. &  43. &   40.  \\ \cline{1-7}
S/$\sqrt{\rm B}$ for 30 fb$^{-1}$
&     2 Tags  &   6.7 &   4.8 &   4.0 &   3.1 &   2.3       \\
\tableline\tableline
%
%
Cross Section (fb) &
SM Higgs     &   40.0 &   32.0 &   25.4 &  19.4 &  13.4  \\
$\sqrt{\rm s~}$ = 14 TeV, p--p & Backgrounds    & 494. &  552. &  508. &  459.
&  449. \\
\tableline
S/$\sqrt{\rm B}$ for 30 fb$^{-1}$ &
     2 Tags  &   9.7 &   7.4 &   6.2 &  5.0 &  3.5  \\
\tableline\tableline
\end{tabular}
\end{center}
\end{table}
%
%
\setcounter{table}{0}
\begin{table}[htb]
\caption{(b) Sensitivity for the
$W(\rightarrow\ell\nu)h(\rightarrow b\bar{b})$ channel}
\begin{center}
\begin{tabular}{||l|c|c|c|c|c|c||}
\tableline\tableline
 & Mass (GeV) &  80 &  90 & 100 & 110 & 120 \\
   &   $\sigma_M$ (GeV)           & 5.5 & 6.0 & 6.5 & 7.0 & 7.4 \\ \cline{1-7}
Cross Section (fb) &
SM Higgs     &    26. &    20. &    15. &    11. &     7.   \\
$\sqrt{\rm s~}$ = 2 TeV, \mbox{p--$\overline{\rm p}$~} &
Backgrounds     &   303. &   303. &   272. &   240. &  207.   \\ \tableline
S/$\sqrt{\rm B}$ for 30 fb$^{-1}$
 &   1 Tag  &   8.1 &   6.1 &   4.8 &   3.8 &   2.7   \\
\tableline\tableline
Cross Section (fb) &
SM Higgs     &    10.5 &  7.9 &  6.0 &  4.4 &  2.9   \\
$\sqrt{\rm s~}$ = 4 TeV, p--p &
Backgrounds     &   42. &   48. &   41. &   36. &   34.   \\ \tableline
S/$\sqrt{\rm B}$ for 30 fb$^{-1}$
 &     2 Tags  &   8.8 &   6.2 &   5.2 &   4.0 &   2.8        \\
\tableline\tableline
Cross Section (fb) &
SM Higgs     &   41.3 &   32.2 &   26.1 &    20.0 &   13.8  \\
$\sqrt{\rm s~}$ = 14 TeV, p--p  &
Backgrounds   &  458. &  518. &  511. &  520. &  540.  \\   \tableline
S/$\sqrt{\rm B}$ for 10 fb$^{-1}$  &
    2 Tags  &  10.5 &   7.8 &   6.4 &  4.9 &   3.3   \\
\tableline\tableline
\end{tabular}
\end{center}
\end{table}
%
%
\setcounter{table}{0}
\begin{table}[htb]
\caption{(c) Sensitivity for the
$Z/W(\rightarrow{jj})h(\rightarrow\tau^{+}\tau^{-})$ channel}
\begin{center}
\begin{tabular}{||l|l|c|c||}
\tableline\tableline
                  & Mass (GeV)  &  110 & 120 \\
\tableline\tableline
$\sqrt{\rm s~}$ = 2 TeV   & Signal Events                & 83.8 & 47.3 \\
\mbox{p--$\overline{\rm p}$~}           & Error on Signal              & 14.5 &
 9.1 \\
\tableline
${\cal L} = 30$ fb$^{-1}$     & Significance                 &  5.8 & 5.2  \\
\tableline\tableline
$\sqrt{\rm s~}$ = 4 TeV   & Signal Events                &  127.5&  72.1   \\
p--p             & Error on Signal              &   30.2&  23.7   \\
\tableline
${\cal L} = 30$ fb$^{-1}$     & Significance                 &  4.2 &   3.0
\\
\tableline\tableline
\end{tabular}
\end{center}
\end{table}
\setcounter{figure}{0}
\begin{figure}[htb]
   \hspace*{1.2in}
%
   \caption{(a) $W(\rightarrow\ell\nu)h(\rightarrow b\bar{b})$ signal.  We show
the case
$M_h \simeq M_Z$ to illustrate the ability to discriminate $h$ from $Z$.}
\end{figure}
\setcounter{figure}{0}
\begin{figure}[htb]
   \hspace*{1.2in}
%
   \caption{(b) $W/Z(\rightarrow jj)h(\rightarrow\tau^{+}\tau^{-}$) signal.
$M_h = 120$ GeV.}
\end{figure}
\setcounter{figure}{1}
\begin{figure}[htb]
   \hspace*{1.2in}
%
   \caption{Combined significance.  See the text for details.}
\end{figure}
%
%
\acknowledgements

S.M. thanks N. Longley, X. Shi, and J. Urheim for informative discussions.
This research is supported by the U.S. DOE.

\end{document}